\documentclass[useAMS,usenatbib,aas_macros,referee]{mn2e}

\usepackage[latin9]{inputenc}
\usepackage{graphicx}

\title[Scaling of oscillation frequencies in rotating stars]{Scaling of oscillation frequencies in rotating stars}
\author[D. Casta\~{n}eda and R. G. Deupree]{D. Casta\~{n}eda$^{1}$\thanks{E-mail:
castaned@ap.smu.ca} and R. G. Deupree$^{1}$\\
$^{1}$Institute of Computational Astrophysics \& Department of Astronomy and Physics, Saint Mary's University, Halifax B3H2S1, Canada\\
}
\begin{document}

\date{Feb 2016}

\pagerange{\pageref{firstpage}--\pageref{lastpage}} \pubyear{2015}

\maketitle

\label{firstpage}

\begin{abstract}
Properties of stars undergoing pulsation such as the well known root-mean-density scaling relation can be useful when trying to match the observed properties of a particular star. It is often assumed that this relation is valid for p mode frequencies in rotating stars. To examine the change in frequency with rotation and mass, we have studied oscillation frequencies of two-dimensional uniformly rotating zero-age main sequence stellar models in the delta Scuti mass range. We identified axisymmetric $p$ and $g$ modes for non-rotating models and then traced them as the rotational velocity was increased. We considered a rotation sequence of ten models for four different masses, with the largest rotation rate being about 200 km s$^{-1}$. The models were required to have the same surface shape between all masses for a given rotation rate. We find that scaling relationships exist among the oscillation frequencies of the same mode for different masses when the models have the same shape. For p modes, this scaling closely follows the period root-mean-density relation found in spherical stars. The $g$ modes also scale between models of the same shape, with the scaling reflecting the change in properties outside the convective core as the stellar mass increases. These scaling relationships can be particularly useful in finding specific stellar models to match the oscillation frequencies of individual stars. We also find that the large separation scales approximately with the root mean density as the rotation rate increases, although the individual mode frequencies do not.

\end{abstract}


\section{Introduction}
It has long been recognized that stellar oscillations provide a unique opportunity to study the interior structure of stars. Thanks to the space based missions MOST (\citealt{Walker2003}), CoRoT (\citealt{Baglin2001}), and Kepler (\citealt{Basri2005}), we now have accurate frequencies for a large number of stars, including intermediate mass main sequence stars such as the $\delta$ Scuti stars (e.g., \citealt{Michel2008,Kjeldsen2010}). While much effort has been expended modeling these stars (e.g., \citealt{Templeton1997,Breger2000,Suarez2014,Ouazzani2015}), a global understanding of the fine structure of these stars still eludes us.

A significant complication for many of these stars is rotation. Very small rotation removes the degeneracy of the non-axisymmetric modes, but only modest rotation is required before the mode splitting is comparable to the large separation, complicating any analysis of an observed frequency spectrum (\citealt{Suarez2010,DeupreeBeslin2010}). Not much more rotation is required before the splitting of the non-axisymmetric modes becomes nonuniform. Rotation also alters the frequencies of the axisymmetric modes (e.g., \citealt{Saio1981,Lignieres2006,Lovekin2008}). Finally, significant rotation deforms the star sufficiently so that the spectral energy distribution (SED) depends on the (usually unknown) inclination between the rotation axis and the observer, so that even placing the star in the HR diagram becomes nontrivial. The picture is brightened somewhat by the advent of optical interferometry, which allows us to determine both the inclination and, to some extent, the amount of rotation through the observed distortion of the stellar surface for a limited number of stars (e.g.,\citealt{VanBelle2001,DomicianodeSouza2003,Aufdenberg2006,Monnier2007,Zhao2009,Che2011}).

For a rotating star the horizontal and radial variables of the perturbations in a linear pulsation analysis are no longer separable as they are for a non-rotating star. In effect this means that the latitudinal component of the eigenfunction can no longer be given by a single Legendre polynomial, but must be regarded as a sum of Legendre polynomials. This has led to two dimensional solutions to the eigenvalue problem (e.g., \citealt{Clement1998,Yoshida2001,Espinosa2004,Lignieres2006,Lovekin2008,Lovekin2009,Ouazzani2012,Ouazzani2015}).

Under the best of circumstances the task, then, is to find a rotating model which has the observed oblateness and oscillation frequencies and which, when viewed at the observed inclination, produces the observed SED. Because the oblateness is tightly coupled to the mass, stage of evolution and the rotation rate, it is unreasonable to expect that just evolving rotating models from the ZAMS to the approximately right place in the HR diagram will satisfy all these conditions. Fortunately, we can simplify the process in various ways.

\cite{Deupree2011a} computed a number of ZAMS models with different mass and rotation rates and found that any two models which have the same surface shape (close to having the same value of $\Omega/\Omega_{\mathrm{crit}}$ but not quite) have the same relative distribution of the effective temperature as a function of latitude.  The effective temperature and effective gravity as a function of latitude and the surface shape are used to compute the SED that would be observed at a given inclination between the rotation axis and the observer (e.g., \citealt{Slettebak1980,Linnell1994,Fremat2005a,Aufdenberg2006,Lovekin2006a}). This suggests that the deduced effective temperature and deduced luminosity, which one would obtain from the SED under the assumption that it originates from a spherical star, might have the same relative distribution as functions of the inclination as well. \cite{Castaneda2014} performed calculations of the SED and showed that this is the case as long as the models are not too dissimilar in structure (much like the period - root mean density relation applies).

Such scaling allows the determination of the surface conditions of a rotating star with relatively little computation. The scaling algorithm assumes that the oblateness and inclination of the star is known from interferometric observations and that there is access to a rotating model with the observed oblateness. Besides the surface properties of this model, the (computed) deduced effective temperatures and luminosities as a function of inclination are required. Also needed are the deduced effective temperature and deduced luminosity at the observed inclination of the star, which can be computed from the observed SED in the same way as they were computed for the model. With this information, the ratio between the surface properties and the deduced properties of the model and the observed star can be calculated, leading to all the surface and intrinsic properties for the rotating star. This is equivalent to placing a non-rotating star in the HR diagram. See \cite{Castaneda2014} for more details regarding the scaling process.

The next step is to compute a model that has the correct shape, surface effective temperature distribution, and total luminosity. Evolution tracks should get one close, and then one converges models with small perturbations to the rotation rate, total mass, and composition distribution to obtain the model. Of course, it is unlikely that this model will have the observed oscillation frequencies, so we must now make corrections to the model to improve the agreement between the observed and model oscillation frequencies. This can be most easily performed if there are scaling relations between the oscillation frequencies of two models, much like the conditions for which the period-root mean density relation for non-rotating stars, to guide us.

Previous studies have been able to relate properties of the oscillation spectra like the large separation with physical properties of the star, particularly the mean density (\citealt{Reese2008,Garcia2013,Garcia2015}) for stars close to the $\delta$-Scuti range of masses. This is can be an important tool when trying to find patterns that can aid the mode identification process. In particular, \cite{Lovekin2009} tested physically meaningful radii that would make the period-mean density relation, $Q=P\sqrt{M/R^{3}}$, constant as the rotation rate increases. Using the radius at $40 \textdegree$ seemed to keep $Q$ constant as a function of the rotation rate, but its selection had no physically meaningful justification. For the models we are considering however, a natural consequence of comparing two stars with different masses but the same surface shape is that by definition their physical volume will be proportional between each other. Additionally, if the range of masses is small enough so that their internal structures are similar, this volume proportionality could potentially relate their oscillation properties directly, just like the period - root mean density scaling relationships for non-rotating stars. One purpose of this paper is to test the validity of this hypothesis. The mean density can be changed by at least three quantities associated with stars; the stellar mass, rotation, and evolution. With our models we can judge the relative importance of each in terms of scaling oscillation frequencies using the mean density

Section 2 presents the procedure followed to obtain the oscillation frequencies of the models considered. In Section 3 we present an overview of the modes found in the frequency range considered, mainly highlighting some of the challenges encountered while each mode was followed as rotation rate increased. Section 4 presents the results of what happens when the oscillation frequencies between two rotating models that have the same surface shape are compared. Section 5 discusses the relative importance of the stellar quantities that can affect the mean density. Finally section 6 contains a review of the results, potential applications as well as the limitations of the relationships found.

\section{CALCULATION OF OSCILLATION FREQUENCIES \label{sec2}}

We follow the same procedure for calculating the oscillation frequencies of rotating stars used by \cite{Lovekin2008}, \cite{Lovekin2009} and \cite{DeupreeBeslin2010}. The method is a two step process: First we use 2D rotating models calculated using the ROTORC code developed by Deupree (\citeyear{Deupree1990}, \citeyear{Deupree1995}). The code solves conservation equations for mass, three components of momentum, energy and hydrogen abundance, as well as Poisson's equation for the gravitational potential. The particular models included in this work are all a subset of the models presented in \cite{Deupree2011a}; they are are zero-age main sequence (ZAMS) models with uniform rotation and solar metallicity with masses  between 1.875 and 2.5 M$_{\odot}$, approximately covering the $\delta$ Scuti mass range. For each mass we picked ten surface equatorial rotational velocities, keeping $R_{\mathrm{pole}}/R_{\mathrm{eq}}$ of the model constant between the different masses at every step in rotational velocity. A summary of the properties of these models is given in Table \ref{tab:rot-models}. It is notable that the ratios between the polar and equatorial effective temperatures are effectively the same for all masses for a given ratio of polar to equatorial surface radius (\citealt{Deupree2011a}). The effective temperatures at each latitude are not determined by von Zeipel's law, but rather are determined by a relation between the local effective temperature and local surface temperature. This makes $\partial \mathrm{log}\,g / \partial \mathrm{log} \, T_{\mathrm{eff}}$  smaller, closer to $0.2$ than to the value of $0.25$ given by von Zeipel's law. This smaller value is similar to that found in previous studies by various means (e.g., \citealt{Monnier2007,Che2011,Espinosa2011,Claret2012}).

\begin{table}

  \centering
  \caption{Rotating models considered \label{tab:rot-models}}
  \begin{tabular}{crrrrc}
  \hline
	& \multicolumn{4}{c}{Mass ($\mathrm{M_{\odot}}$)} &\\
    & $1.875$ & $2.000$ & $2.250$ & $2.500$ &\\
     Shape ($R_{\mathrm{pole}}/R_{\mathrm{eq}}$)& \multicolumn{4}{c}{Velocities ($\mathrm{km\,s^{-1}}$)} & $T_{\mathrm{eff,pole}}/T_{\mathrm{eff,eq}}$\\

 \hline
 1.000 & 0.0 & 0.0 & 0.0 & 0.0 & 1.000 \\
 0.997 & 35.0 & 36.0 & 36.0 & 37.5 & 1.003 \\
 0.991 & 62.0 & 63.0 & 65.0 & 67.0 & 1.009 \\
 0.985 & 83.0 & 84.0 & 87.0 & 89.0 & 1.015 \\
 0.976 & 105.0 & 106.0 & 109.0 & 111.0 & 1.024 \\
 0.966 & 125.0 & 127.0 & 131.0 & 134.0 & 1.034 \\
 0.954 & 146.0 & 148.0 & 152.0 & 156.0 & 1.047 \\
 0.940 & 165.0 & 168.0 & 173.0 & 178.0 & 1.064 \\
 0.925 & 187.0 & 190.0 & 195.0 & 200.0 & 1.079 \\
 0.907 & 207.0 & 211.0 & 217.0 & 222.0 & 1.098 \\

\hline
\end{tabular}

\end{table}

We calculated pulsation frequencies using the linear, adiabatic, non-radial pulsation code developed by \cite{Clement1998}. This code expresses the eigenfunctions of each mode as a sum of up to eight spherical harmonics with a radial resolution of 500 zones. In particular, for this study we calculated modes with $\ell$ values up to $15$ (even and odd parity modes can be computed separately). The primary motivation for the number of basis functions is to obtain the frequencies of the low $\ell$ modes accurately,  not to compute the highest $\ell$ modes. For this reason, we choose to include in our analysis only values of $\ell$ between $0$ and $6$. Similarly, we focus our discussion on the radial orders of $p$ modes up to $n=8$ in the non-rotating case because of our limited resolution near the model surface. We then follow these specific modes as a function of rotation, although we do take into account avoided crossings between the modes we are interested in and the higher $\ell$ modes. As was found by previous studies using this code (\citealt{Lovekin2008, DeupreeBeslin2010}), we find that 8 spherical harmonics can yield accurate eigenfrequencies for selected modes and rotational velocities considered. Specifically we found that the difference in computed frequencies of modes using 6 and 8 basis functions is on average on the order of a few thousands of a percent for the lowest rotating models and about a tenth of a percent for the highest rotating models. These frequency differences indicate that at least for the modes selected, the number of basis functions is adequate.

For rapidly rotating stars the process of classification of modes is more problematic than for non-rotating stars. The major reason is because the usual latitudinal quantum number, $\ell$, is no longer sufficient to specify the latitudinal variation of the eigenfunction. Similar to other studies, (e.g., \citealt{Lignieres2006, Lovekin2008,Lignieres2009, DeupreeBeslin2010}) we address this problem by characterizing each mode with the value of $\ell$ associated with the mode in the non-rotating model, $\ell_{0}$. This tracing process, although useful, becomes progressively more difficult for more rapidly rotating models. For each model we identified the same combination of $g$ and $p$ modes as well as several radial modes. In total, we computed 180 modes with even parity and 189 modes with odd parity. Table \ref{tab:modes} presents the non-rotating modes identified. Of these, the number of modes which match the analysis criteria described above is 78. Under the classification scheme developed by \cite{Lignieres2006,Lignieres2009,Reese2009} in which modes for rapidly rotating stars can be separated as either island, whispering gallery or chaotic modes, we expect our selected low degree modes to be mainly island modes.

\begin{table}
  \centering
  \caption{Oscillation modes calculated for each model. The value of
  $n$ denotes the number of radial nodes found in mode for the non-rotating case \label{tab:modes}}
  \begin{tabular}{ccc}
  \hline
    $\ell$ & $n$-range ($g$ mode) & $n$-range ($f$ and $p$ mode)   \\

 \hline
 0 & &0-15 \\
 1 & 1 & 1-16 \\
 2 & 1-2 & 0-14 \\
 3 & 1-3 & 0-15 \\
 4 & 1-4 & 0-15 \\
 5 & 1-5 & 0-15 \\
 6 & 1-7 & 0-15 \\
 7 & 1-8 & 0-14 \\
 8 & 1-9 & 0-13 \\
 9 & 1-10 & 0-13 \\
 10 & 1-11 & 0-13 \\
 11 & 1-13 & 0-13 \\
 12 & 1-13 & 0-12 \\
 13 & 1-15 & 0-12 \\
 14 & 1-16 & 0-12 \\
 15 & 1-18 & 0-12 \\

\hline
\end{tabular}
\end{table}

\section{VARIATION OF MODE FREQUENCIES WITH ROTATION \label{sec3}}

\begin{figure}
  \centering  
  \includegraphics[width=0.8\columnwidth]{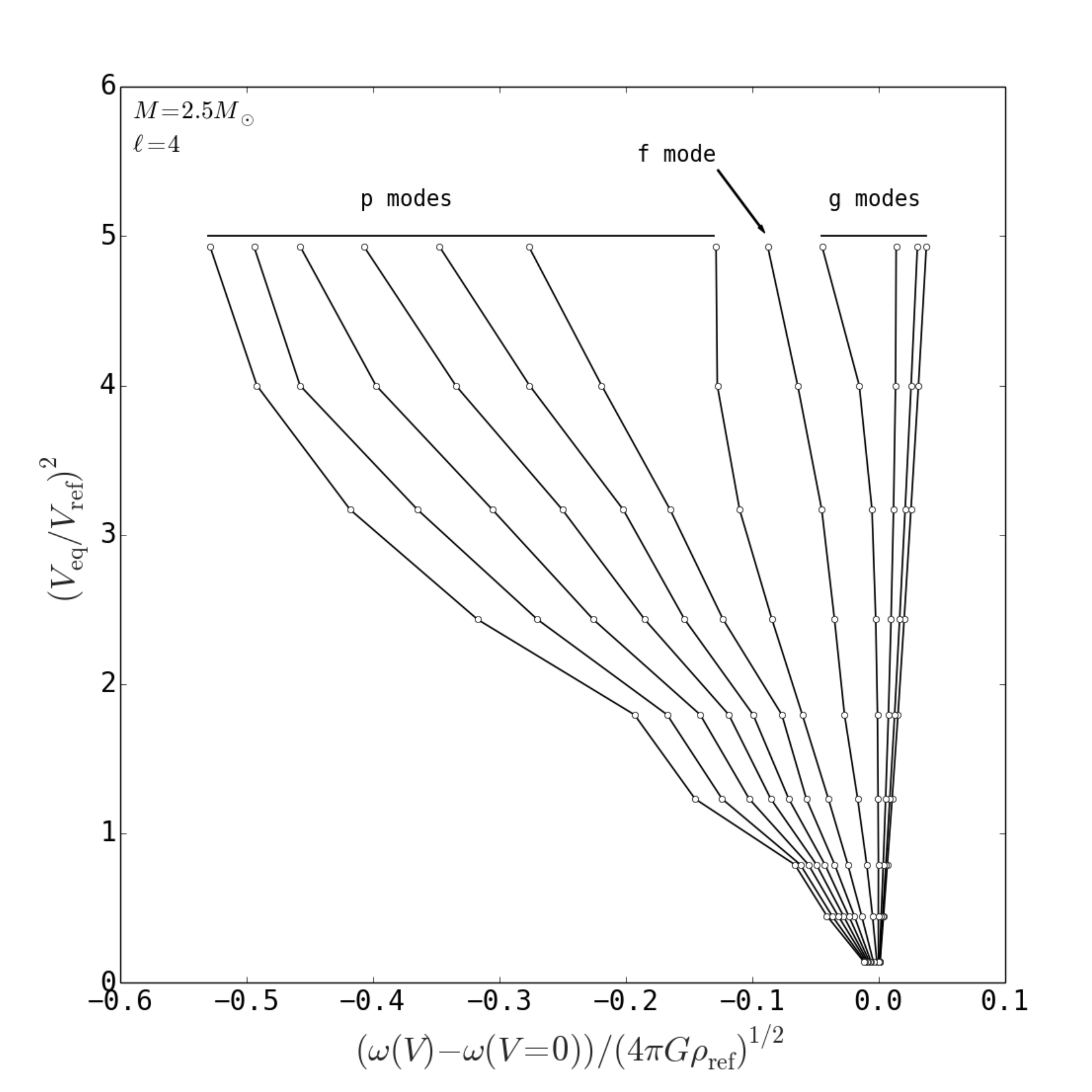}
  \caption{Rotational evolution of $\ell_{0}=4$ mode for $M=2.5\mathrm{M_{\odot}}$. Frequencies are scaled by the factor $(4\pi G \rho_{\mathrm{ref}})^{1/2}$, where $\rho_{\mathrm{ref}}=1\mathrm{g \, cm^{-3}}$, to keep their values at around unity across models and $V_{\mathrm{ref}} = 100$ km s$^{-1}$. Some of the changes in direction, most evidently for the first $p$ mode, are caused by interactions with other $\ell$ modes as well as changes of properties of each mode as rotation is higher. Radial order increases to the left for the $p$ modes and to the right for the $g$ modes}
  \label{fig:rot-evol1}
\end{figure}

Significant research has been performed to determine the behavior of oscillation frequencies with rotation. Some studies have used the ``traditional'' approximation in which the horizontal component of the angular momentum is neglected in the linearized momentum equations (e.g., \citealt{Lee1986,Lee1997,Townsend2003}). Two other prominent methods are perturbation techniques (e.g., \citealt{Saio1981,Dziembowski1992,Soufi1998,Suarez2010}) and, in principle the more reliable, numerical integrations with the horizontal variations assumed to be given by a sum of spherical harmonics (e.g., \citealt{Clement1998,Lignieres2006,Lovekin2008,Ballot2010}). Other studies have also explored the problem using ray theory principles (e.g, \citealt{Dintrans2000,Lignieres2009,Pasek2012,Prat2016}) to compute oscillation frequencies in the asymptotic regime. Here we use Clement's radial finite difference integration approach in the radial direction. Furthermore, we will consider only axisymmetric modes. We shall distinguish different latitudinal modes by ``$\ell$'', even though the modes are actually linear combinations of different $\ell$'s.  When we use a specific value of $\ell$, we intend this to mean the value to which a given mode could be traced back to at zero rotation. Figure \ref{fig:rot-evol1} shows the general evolution of oscillation frequencies with rotation. The $p$ mode frequencies decrease, mostly because the stellar volume increases, and hence the mean density decreases, with rotation. The $g$ modes, on the other hand, generally increase in frequency as the rotation rate increases (e.g., \citealt{Ballot2010}), but by a much smaller amount. This figure also shows a subset of low radial order $g$ modes that increase in frequency with increasing rotation at low rotation, but then begin to decrease with frequency as the rotation increases further.

To discuss these and other features, we divide the frequency range covered into four intervals: 1) low frequency which is defined as having only $g$ modes, 2) an intermediate frequency region which includes both $g$ and $p$ modes, 3) a higher frequency region which includes only $p$ modes, but at frequencies sufficiently low that there are still sizable variations in the frequency difference between radial quantum numbers $n$ and $n$+1 for different latitudinal modes, and 4) the high frequency regime in which the large separation is essentially the same for all latitudinal modes. We shall address each of these frequency regions in turn.

\begin{figure}
  \centering  
  \includegraphics[width=0.8\columnwidth]{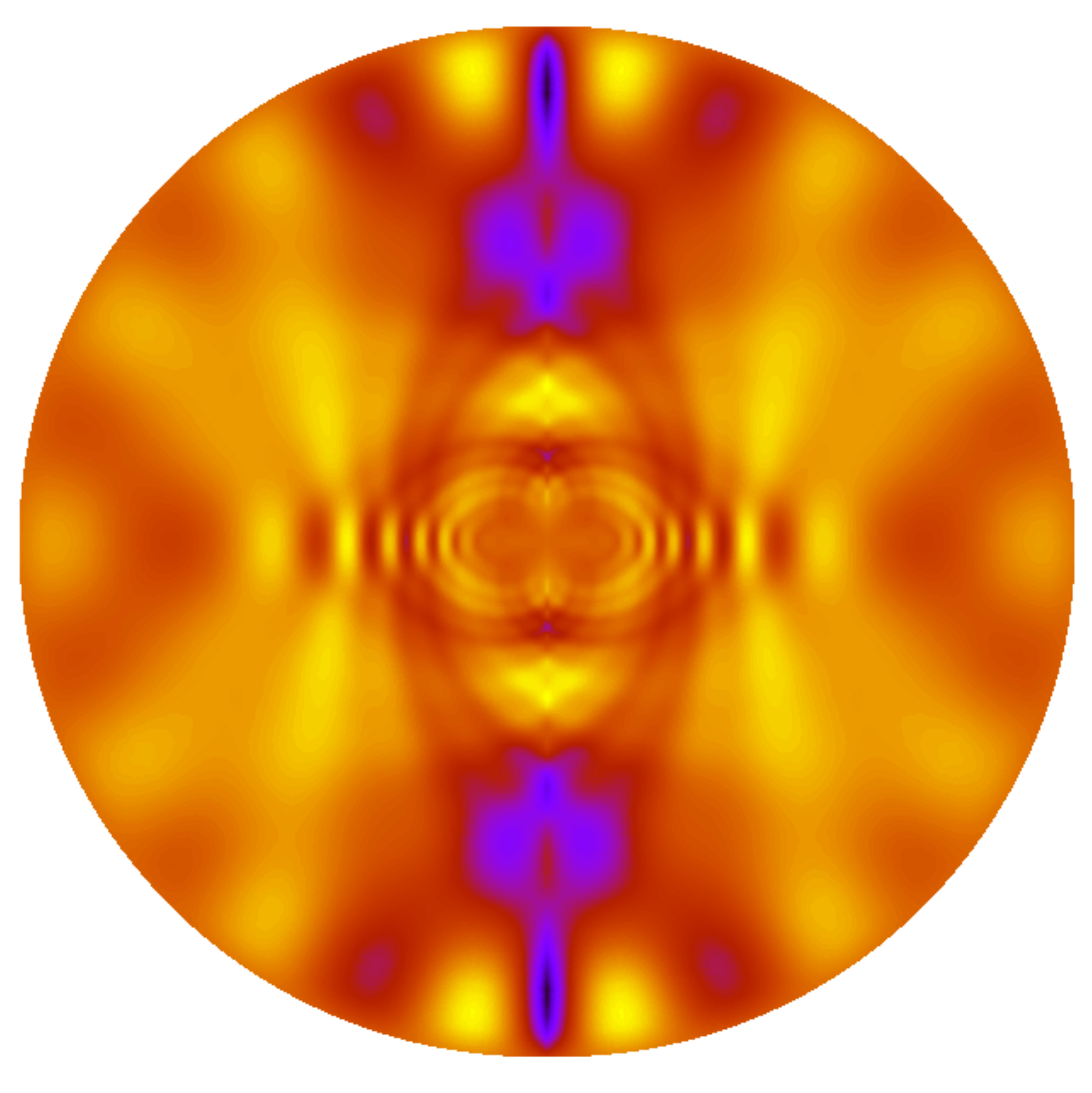}
  \caption{Meridional cross section of the pressure perturbation of a $g$ mode ($\ell=3$, \, $n=-3$) showing the rosette-like structure. A colour version of this figure is available online}
  \label{fig:rosette}
\end{figure}

The low frequency $g$ modes in our set are perturbed much less by rotation than the $p$ modes for the uniformly rotating models. This is because the $g$ mode frequencies are most sensitive to the region  just exterior to the convective core boundary while the rotational effects on the model structure are substantial only near the surface. Both the frequencies and the eigenfunctions in this interval are modestly perturbed, making mode identification relatively easy. The change in frequency from the non-rotating value for a given rotation rate is larger at lower frequencies, increasing by about 40\% as the frequency is decreased by about a third. There seems to be little dependence with $\ell$. This means that avoided crossings are rare, but they are not entirely absent. It would appear that a necessary condition is that the g mode frequencies are already quite close in the non-rotating model. Interestingly, previous studies (e.g., \citealt{Ballot2012,Takata2013}) have found that some of these $g$ modes with frequencies close to each other show particular rosette pattern in their internal structure. It is still not clear whether these modes are excited in real stars but we can confirm that some rosette modes are part of our data set. Figure \ref{fig:rosette} shows an example of one case found in our sample, in which a $g$ mode with $\ell=3$ and $n=3$. The frequency evolution of these modes, however, does not seem to differ from the non-rosette modes, making them only interesting in their internal structure. Finally, we note that even our lowest frequency modes are somewhat above the sub-inertial range (thereby limiting the impact of the Coriolis force) and that the difference between their pulsation frequencies and the corresponding non-rotating values increases as the square of the rotation rate, reflective of the centrifugal force.

\begin{figure}
  \centering  
  \includegraphics[width=0.8\columnwidth]{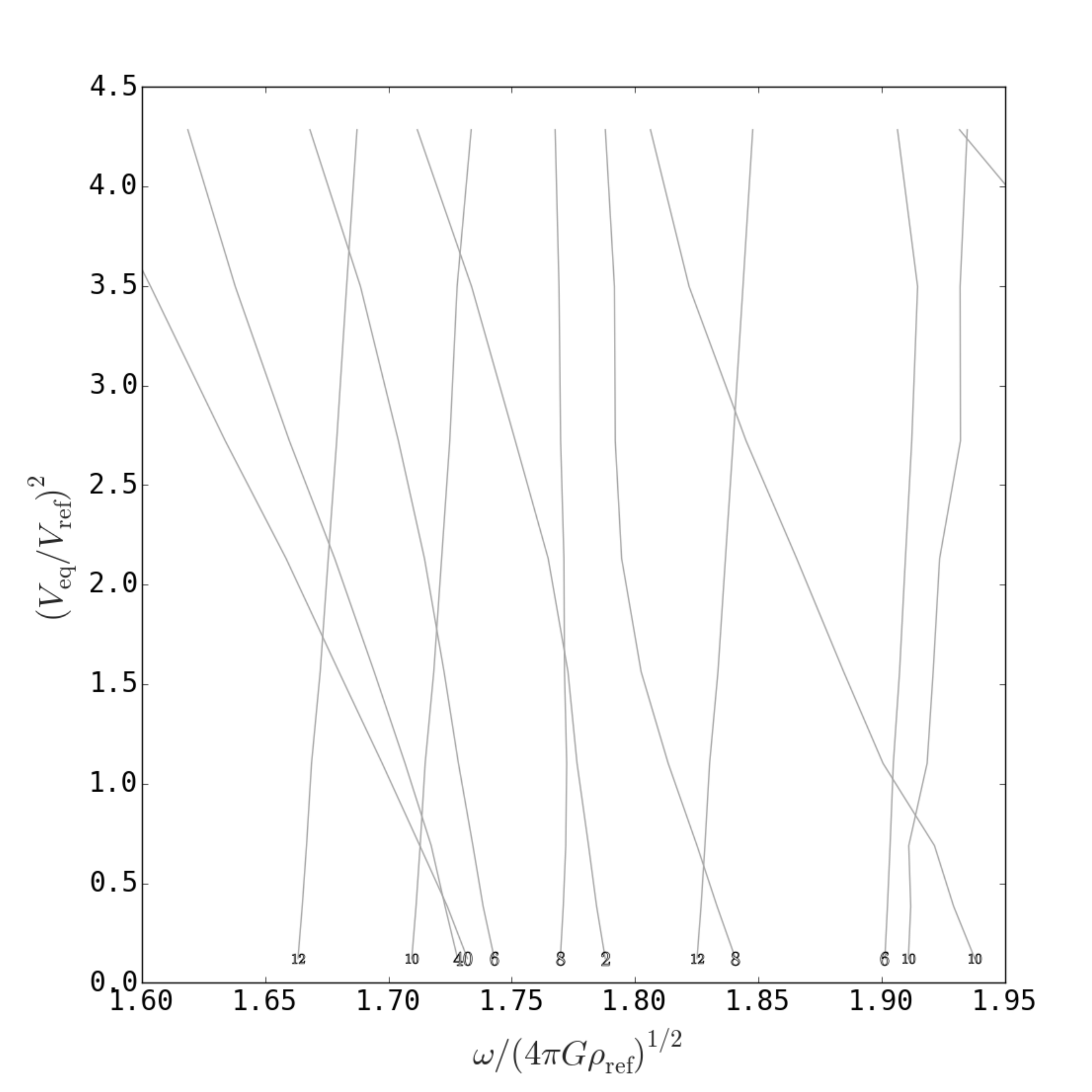}
  \caption{Rotational evolution of $\ell$'s with even parity calculated for the $1.875M_{\odot}$ models. $V_{\mathrm{ref}}$ and $\rho_{\mathrm{ref}}$ are as defined in Figure 1. We have labeled each line with its corresponding ``initial'' $\ell$. There are a few cases of interaction between $p$ modes whose frequency decreases as rotation increases and $g$ modes whose frequency evolution behavior is the opposite. It is important to note that the lines do not cross, they correspond to avoided crossings. Modes in this region are the hardest to follow because in some cases modes can interact with each other two or three times before reaching the most rapidly rotating models included.}
  \label{fig:mixed-modes}
\end{figure}

The region in which there are both $p$ and $g$ modes is the most complex. This region is relatively small in frequency width (the width increases as the rotation rate increases), but the mode density is relatively high and the different signs of frequency change for the $p$ and $g$ modes with increasing rotation make near resonances and avoided crossings comparatively common. We present an example of avoided crossings in Figure \ref{fig:mixed-modes} and also note from Figure \ref{fig:rot-evol1} that the first $p$ mode shows a change in its frequency evolution trend caused by one of these avoided crossings. Avoided crossings are relatively more common at higher $\ell$ because higher $\ell$ $g$ modes for a given $n$ have higher frequencies. Similarly, they are more common at lower $n$ because lower $n$ $g$ modes have higher frequencies.

It is also in this interval that we see the phenomenon mentioned above where a given $g$ mode frequency increases with increasing rotation up to a point and then decreases with further rotation (see Fig. \ref{fig:rot-evol1}). These are mixed modes that have nodes in the interior where the $g$ mode nodes are expected and also much nearer the surface where $p$ mode nodes occur (e.g., \citealt{Scuflaire1974,Osaki1975}). The frequency may actually begin to increase with increasing rotation before the node near the surface appears. This is similarly indicated in phase diagrams, such as defined by \cite{Scuflaire1974} and \cite{Osaki1975}, in which the latitudinal displacement or, equivalently,  $\frac{P^{'}}{\rho}+\psi^{'}$ (the primes denote Eulerian perturbations) is plotted against the radial displacement. The latitudinal displacement and the pressure-gravity perturbation are functionally equivalent in a non-rotating model, but not in a rotating one. We use the Eulerian pressure-gravity perturbation in part for convenience and in part because of the similarity of the latitudinal variation of this perturbation and the radial displacement. The phase diagrams are computed along radial lines (i.e., at a given latitude). At low rotation rates the phase diagrams at different latitudes show essentially the same clockwise or counterclockwise behavior, but the values of neither variable are the same. This is most noticeable at the surface, where the final point on the diagram may end in different quadrants for different latitudes. At higher rotation rates there may be different numbers of nodes at different latitudes, and in the phase diagram this may appear as loops which do not quite cross one or the other axes at some latitudes but do at other latitudes or as a direction reversal which crosses an axis for some latitudes but stops short of crossing at others. The relationship between the number of nodes and a specific mode identification becomes more murky at the highest rotation rates we use, although the motion in the phase diagram is still clockwise for g mode behavior and counterclockwise for $p$ mode behavior with the type of mode behavior determined by where the nodes are in radius. The change of the phase diagram with rotation for a mode that will become mixed as the rotation rate increases is as follows. As the radius increases, the curve proceeds in a clockwise direction for $g$ modes, but for these $g$ modes that show a frequency decrease with increasing rotation, the motion becomes counterclockwise at sufficiently large fractional radius, although it does not continue far enough to produce a node near the surface. Because the effect increases as the rotation rate increases, one expects that $p$ nodes will appear at yet more rapid rotation. We do see f modes that are mixed in these non-rotating ZAMS models, and adding rotation eventually leads to eigenfunctions in which the node pattern is more pronounced. Rotation also extends mixed modes to some low radial order $g$ modes. There may be an avoided crossing before this mixed original $g$ mode occurs.

\begin{figure}
\begin{minipage}[t]{0.49\columnwidth}%
\begin{tabular}{c}
\includegraphics[scale=0.3]{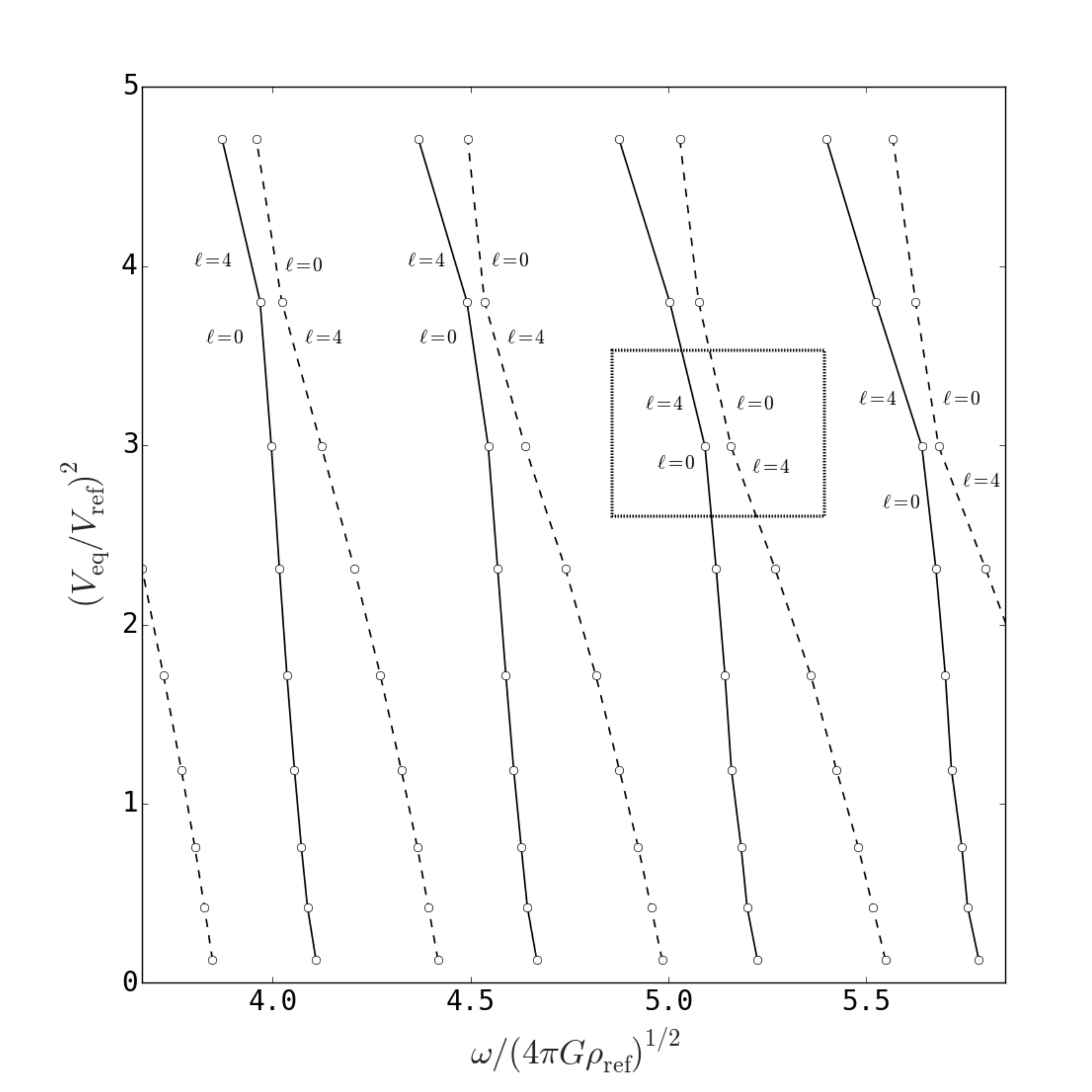}\tabularnewline
\tabularnewline
\end{tabular}
\end{minipage}
\begin{minipage}[t]{0.49\columnwidth}%
\begin{tabular}{ll}
\includegraphics[scale=0.15]{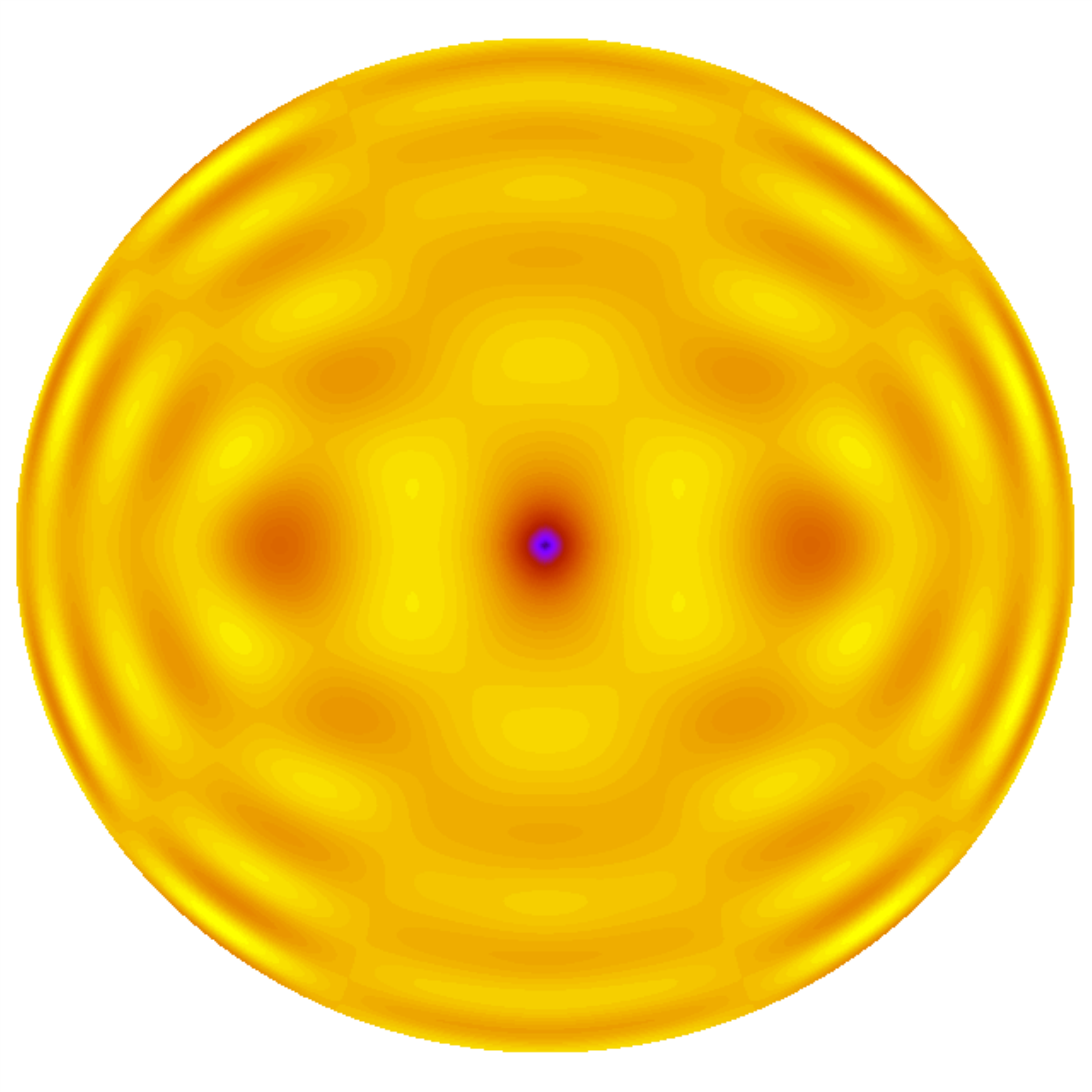} & \includegraphics[scale=0.15]{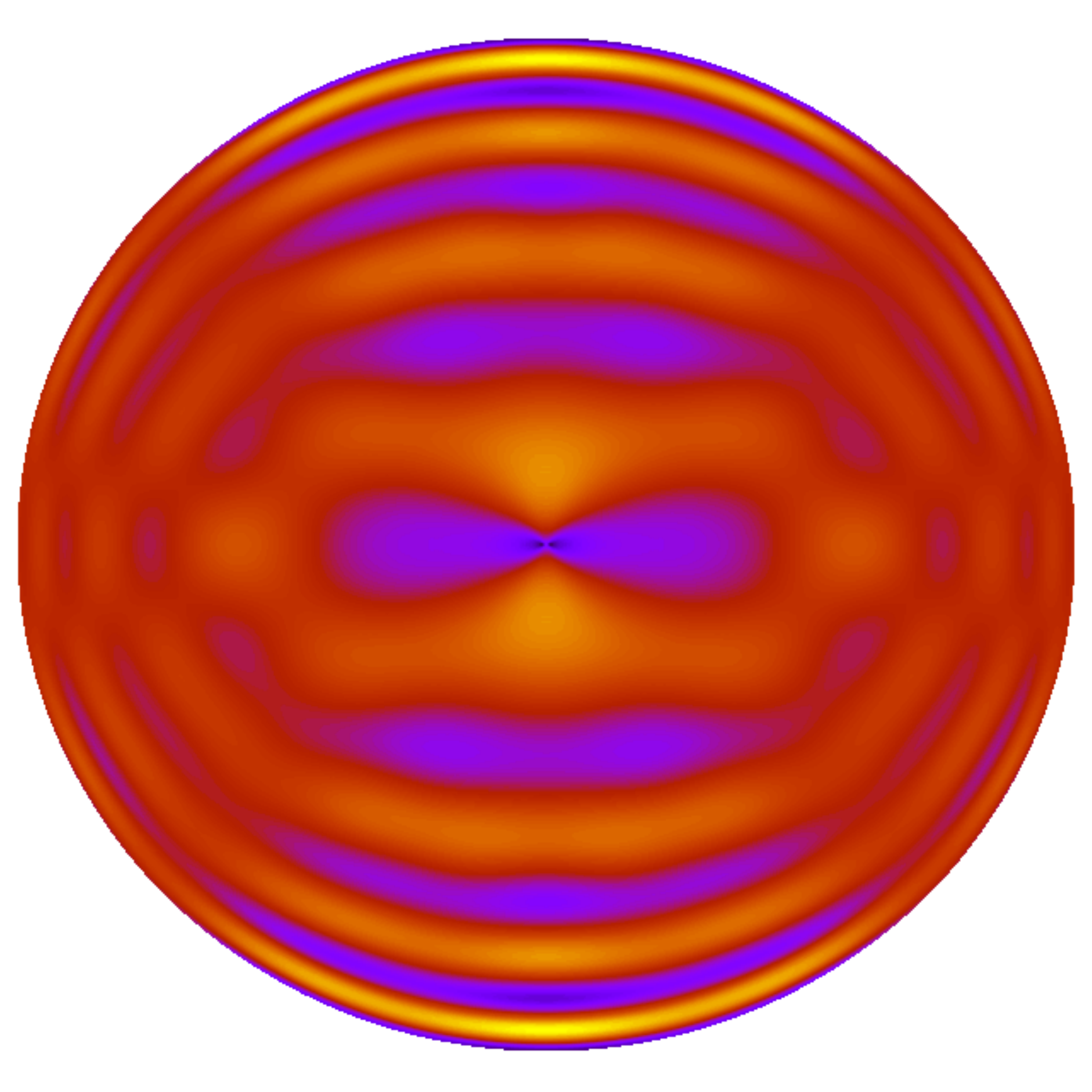}\tabularnewline
\includegraphics[scale=0.15]{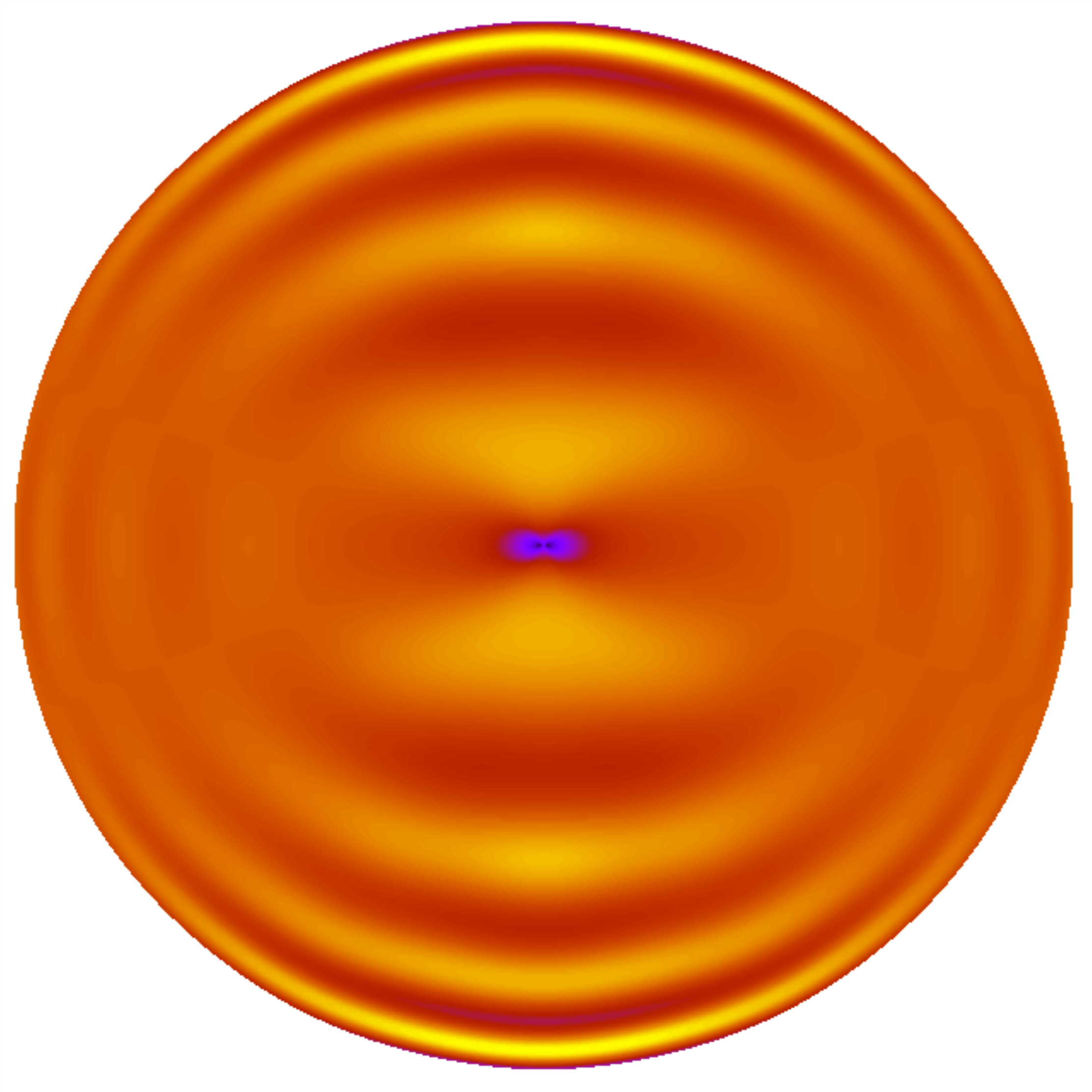} & \includegraphics[scale=0.15]{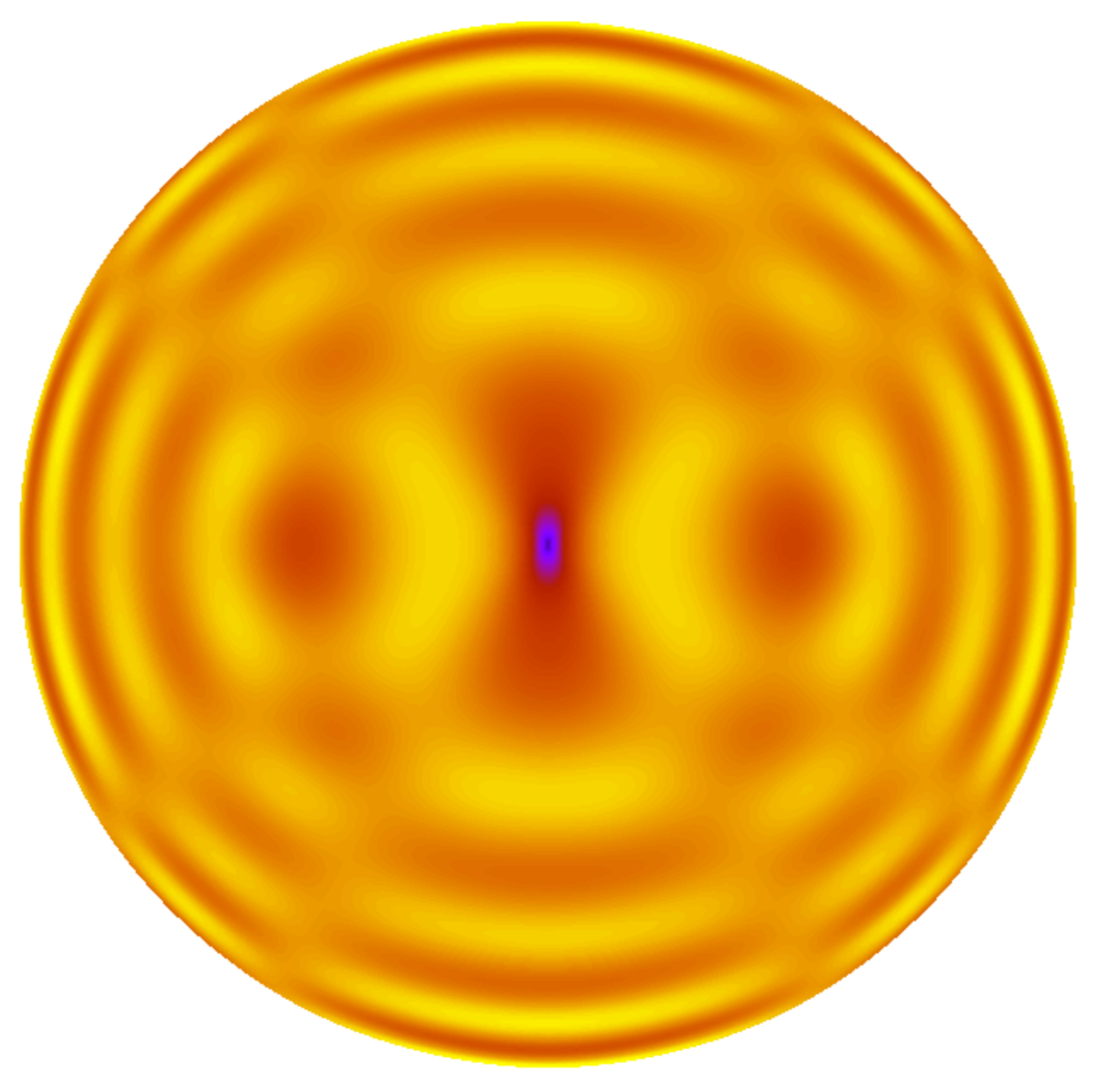}\tabularnewline
\end{tabular}
\end{minipage}

\caption{Mode interaction between modes with $\ell_{0}=0$ and $\ell_{0}=4$ for models of mass $M=2.25\mathrm{M_{\odot}}$. $V_{\mathrm{ref}}$ and $\rho_{\mathrm{ref}}$ are as defined in Figure 1. It is possible to see the change in ``label'' required to match the exchange of surface latitudinal variation for each mode. The 2D contour plots on the right show the before and after of the meridional cross section of the pressure perturbations of the interacting modes enclosed in the box. In the case of the pressure perturbations, the lower frequency mode is on the left.\label{fig:avoided}}
\end{figure}

Avoided crossings also occur in the third frequency region where there are no $g$ modes. The usual case is a low latitudinal order $p$ mode decreasing its frequency with increasing rotation reaching the frequency of a higher latitudinal order $p$ mode that is decreasing its frequency more slowly with increasing rotation. Some easily identifiable crossing in our models appear to involve  $\ell = 0,\,4,\,8$, although other cases certainly occur, in agreement with the crossing pattern found by \cite{Lignieres2006}. We show one such progression in Figure \ref{fig:avoided}. The internal structure illustrations on the right show how it is still fairly easy to differentiate between the two modes involved after the interaction.

The final frequency interval is where $n$ is sufficiently high that the large separation is virtually the same for all $\ell$ values. Thus, one has as many modes as one has spherical harmonics within a frequency interval given by the large separation. The large separation decreases very slightly with increasing rotation, as was found by \cite{Lignieres2006} and by \cite{Lovekin2008}. Another feature is that the eigenfunction for a given $\ell$ no longer changes very much from one value of $n$ to the next for a given latitudinal mode. It was hoped that this frequency regime would be useful in mode identification, but the placement of modes within the large separation, including the ordering, may be different from one rotation rate to the next. For higher frequency modes techniques like the application of ray tracing methods have been successful at finding insights into the frequency patterns that could help their classification (e.g., \citealt{Lignieres2009,Pasek2012}). These methods, however, do not replace the need for highly resolved two dimensional calculations that provide more exact eigenfrequencies and eigenfunctions.

It is interesting to note that the trends for these intermediate mass ZAMS models are similar to those of the 10 M$_{\odot}$ model of \cite{Lovekin2008} and the rotating polytropic models of \cite{Lignieres2006}.

\section{VARIATION OF MODE FREQUENCIES WITH MASS \label{sec4}}

The period-root mean density relation has long been a tool for scaling the oscillation frequencies or, alternatively, estimating the mean density for a model or star for which there is an object sufficiently similar in properties to make the comparison. It has often been tacitly assumed to remain useful for rotating stars. Strictly speaking, one can only expect it to be applicable to $p$ modes, and our discussion about the different behavior of $p$ mode and $g$ mode frequencies with rotation suggests that the period-root mean density law does not apply for $g$ modes. Here we will determine the applicability of the period-root mean density law for $p$ modes of rotating stars and examine the scaling of $g$ modes.

\begin{figure}
  \centering  
  \includegraphics[width=0.49\columnwidth]{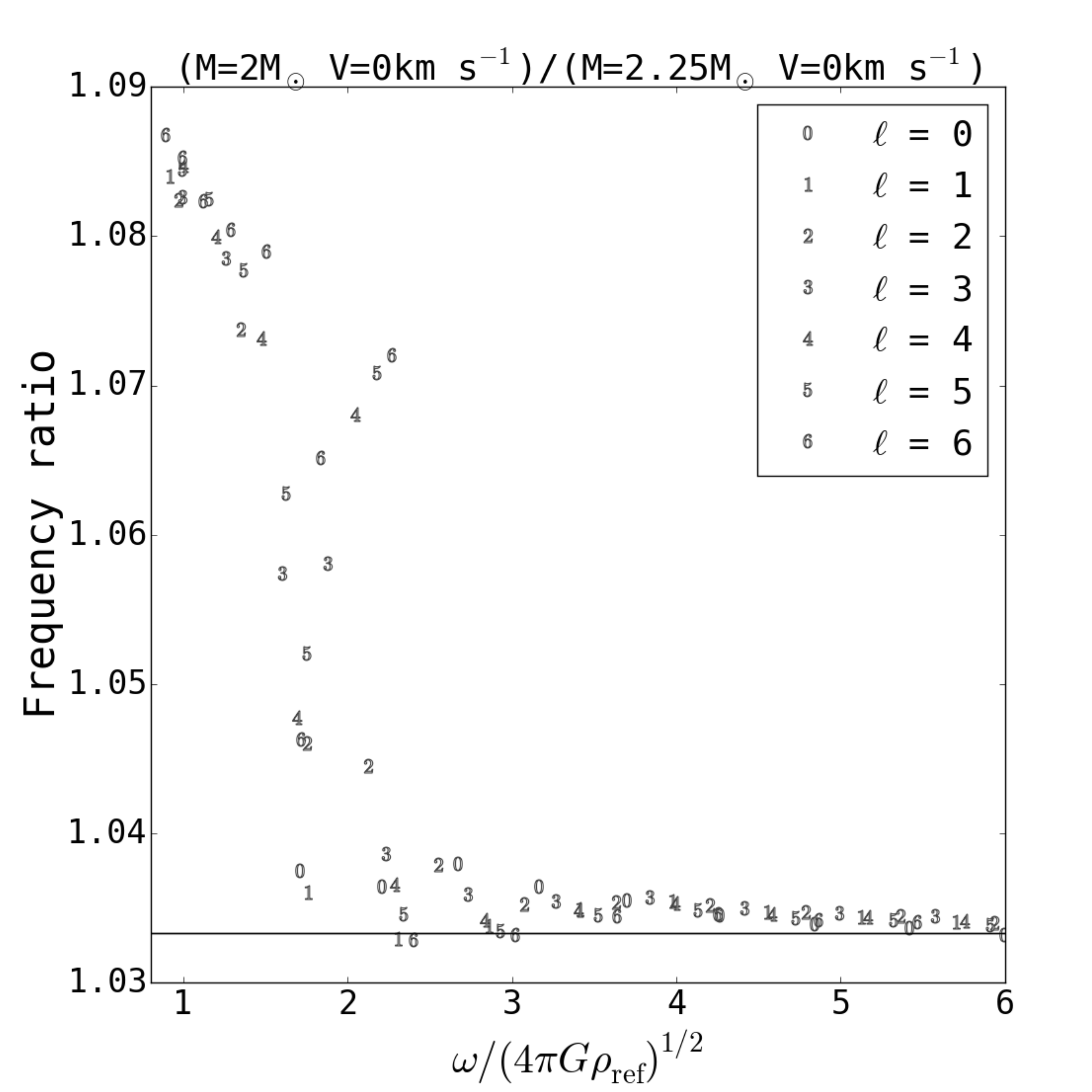}
  \includegraphics[width=0.49\columnwidth]{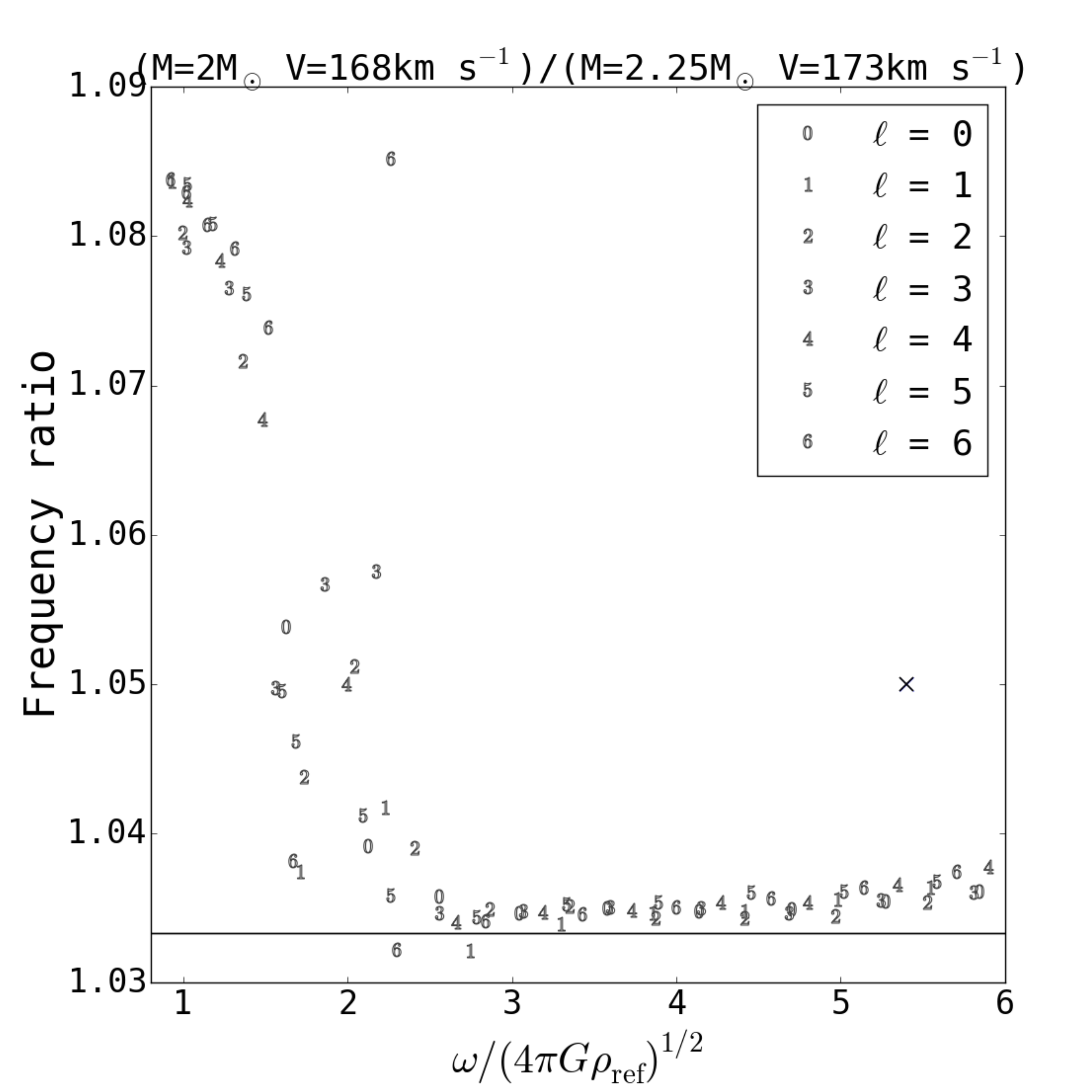}
  \caption{Ratio between frequencies of stellar models with the same shape. Left presents the non-rotating case. Right shows the same ratio for two rapidly rotating models. The black horizontal line indicates the ratio between the root-mean densities of both models. The cross shown on the figure of the right indicates what the ratio between modes that have undergone an avoided crossing for one mass but not the other. See the text for more details.}
  \label{fig:freq-ratio}
\end{figure}

Figure \ref{fig:freq-ratio} presents the frequency ratio of the same modes for the ZAMS models of 2 and 2.25 M$_\odot$. Figure \ref{fig:freq-ratio} (left) shows the results for non-rotating models and Figure \ref{fig:freq-ratio} (right) shows the same calculations for moderately rapidly rotating models. In each case the ratio of the two root mean densities is shown by the horizontal line. It should be noted that the two rotating models having the same surface shape means that the scaling of the stellar volumes will be exact and given by the cube of the ratio of the equatorial radii. It is clear that the effect of rotation on the $p$ mode ratios is rather small and that the period-root mean density is still a useful indicator of how to change the model to obtain desired period changes.

It is also clear that the $g$ modes do not follow this rule, although the ratios also do not depend very much on rotation. We can perhaps obtain some insight into how this $g$ mode ratio comes about by applying equation equation (A13) from \cite{Tassoul1980} to the stable $g$ modes with radial nodes located between the outer boundary of the convective core and the surface ionization regions
\begin{equation}
\label{eq:tassoulA13}
\omega=\frac{(\ell(\ell+1))^{1/2}\int_{r1}^{r2}N\frac{dr}{r}}{(n+1/2)\pi-\theta_{1}-\theta_{2}}
\end{equation}
Here $N$ is the Brunt-Vaisala frequency, and the thetas are phase shifts which occur if either $N$ or the density or their derivatives are discontinuous at the limits of the integration. Because $N$ is taken to be identically zero in the convective core, its first derivative may be discontinuous at the convective core boundary. The integral is taken between two radial locations which are consecutive zeros of $N$, in this case the convective core boundary and the lower boundary of the second helium ionization region. Assuming that we have successfully identified the same mode and the phase shifts are the same for each of the two models, the ratio of the two mode frequencies will be

\begin{equation}
\label{eq:tassoulA13_ratio}
\frac{\omega_{a}}{\omega_{b}}=\frac{\int_{r1a}^{r2a}N\frac{dr}{r}}{\int_{r1b}^{r2b}N\frac{dr}{r}}
\end{equation}

Performing the integrations with the two non-rotating models in Figure \ref{fig:freq-ratio}, we obtain a ratio of $1.088$, clearly close to the upper limit for the ratios of the linear, adiabatic frequencies. For all masses, the largest contributions to the integrals come from the region exterior, but relatively close, to the convective core. This is true both because $N$ is relatively large and $r$ comparatively small there. However, the contributions all the way out to the bottom of the second helium ionization region are not entirely negligible, even if less prominent than the deeper regions.

We also found that as rotation rate increased some of most interesting individual mode ratios were found in the frequency region that contains both $p$ and $g$ modes. For the non-rotating models, the modes with erratic ratio behavior with rotation are either the $f$ mode or low radial order $g$ modes, with no particular preference in latitudinal order. However, mode ratios in this region become significantly more complex with rotation. The ratios of some of these modes change very little with rotation, while the ratios for others change significantly. While no one feature accounts for the modes whose ratio changes significantly, modes bumped, modes with near resonances, and mixed modes appear to be more likely to display this behavior than others. The order at which modes appear in the non-rotating case in this region also seems to be different for every mass and can can also account for the erratic frequency ratio behavior. For the non-rotating cases about 12-18 modes out of the about 140 modes in this region, including all $\ell$ values, change places. Most of the modes involved in this ordering change, however, are modes with $\ell > 8$ and not taken into account for the ratio analysis.

We also see from Figure \ref{fig:freq-ratio} that the ratios do not change significantly as rotation is included (except as we have indicated), at least up to the rotation rates presented here. We suggest that one may use the period-root mean density relationship for $p$ modes in rotating models with the same reliability and constraints as one uses it in non-rotating models. The lower frequency $g$ modes also scale predictably; it is only the modes in this intermediate region which are less predictable.

Finally note that in Figure \ref{fig:freq-ratio} we have followed the mode frequency evolution, not the shape of the latitudinal variation of the eigenfunction. This makes no difference in terms of the computed ratio unless one has the case where a particular mode has undergone an avoided crossing for one mass but not the other. These are rare occurrences. Following the shape in this case changes the frequency ratio for both modes. We show the approximate shift by the cross in Figure \ref{fig:freq-ratio} (right). In terms of increasing rotation rate, the ratios look normal until we reach the rotation rate where the avoided crossing has occurred in one model but not the other. Once the avoided crossing has occurred in both models the ratios look normal again. 

\begin{figure}
  \centering  
  \includegraphics[width=0.99\columnwidth]{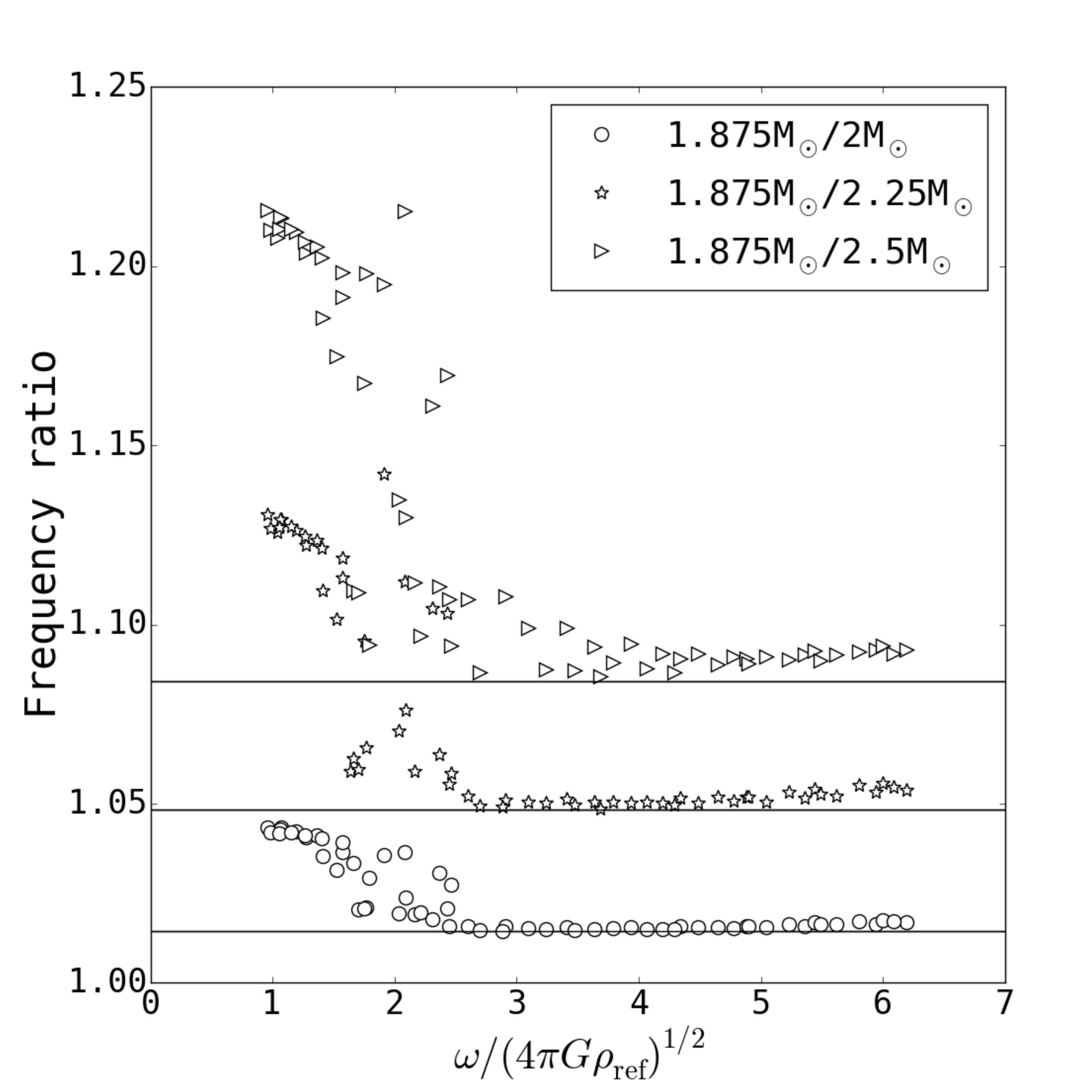}
  \caption{General scaling behavior of the frequency ratios between the computed frequencies of the M=1.875$M_{\odot}$ and the other masses considered for the eighth most rapid rotating models in our set with velocities around 170 km s$^{-1}$. Modes with $\ell$ values from 0 through 6 were included. It can be seen how $g$ and $p$ modes scale differently. The horizontal lines correspond to the theoretical ratio of the root-mean density relation for each pair of masses. The $p$ modes fall close to these lines in every case, but depart from the lines progressively more as the relative difference between the two masses increases.}
  \label{fig:ratio_avg}
\end{figure}

Figure \ref{fig:ratio_avg} shows an overview of how the scaling of frequencies look for all masses considered using the eighth most rapid rotating models in our set with velocities around 170 km s$^{-1}$. It can be seen how the $g$ modes scale differently than the $p$ modes in every case. To give a general idea of how the ratios of the $p$ modes scale for the different mass combinations as well as to see how each $\ell$ behaves, the average ratio between each acoustic mode with $n>1$ and the theoretical root-mean density ratio was calculated. Most ratios fall within 0.1\% and 1\% from the theoretical ratio, where the larger differences correspond to the ratios between models with the biggest mass difference.

\section{DEPENDENCIES OF THE $\Delta \nu -\sqrt{ \overline{\rho }}$}

Several studies have been undertaken to examine the utility of the $\Delta \nu -\sqrt{ \overline{\rho }}$ relation for  $\delta$ Scuti variables (e.g., \citealt{Reese2008,Suarez2014,Garcia2015}). Here $\Delta \nu$ is taken to refer to frequency differences with $\Delta n=1$ and $\Delta \ell =0$ for relatively small $n$ instead of in the limit of large $n$. We discuss the relative importance of three factors that affect this relationship: evolutionary state, rotation, and mass for models with the same rotational shape.

The largest variation in the mean density comes from the evolution of the star along the main sequence, dropping by a little less than a factor of six from the ZAMS to the end of core hydrogen burning for a 2$M_{\odot}$ non-rotating model. In contrast, the change in mean density between our 2$M_{\odot}$ non-rotating model and a 2$M_{\odot}$ model rotating near critical rotation (\citealt{Deupree2011a}) is less than a factor of two.  The change between the mean density of the non-rotating model and the highest rotation rate we consider in this paper is only about 10\%. We obtain the same mean density as that for our most rapidly rotating model for an evolved model with $X_{c}$ = 0.64 (the ZAMS models have $X_{c}$ = 0.70). Finally, we note that the mean density changes by about 17\% between our 1.875$M_{\odot}$ and 2.5$M_{\odot}$ ZAMS models (the ratio is the same for all rotation rates because of the way the models were constructed).

\begin{figure}
  \centering  
  \includegraphics[width=0.99\columnwidth]{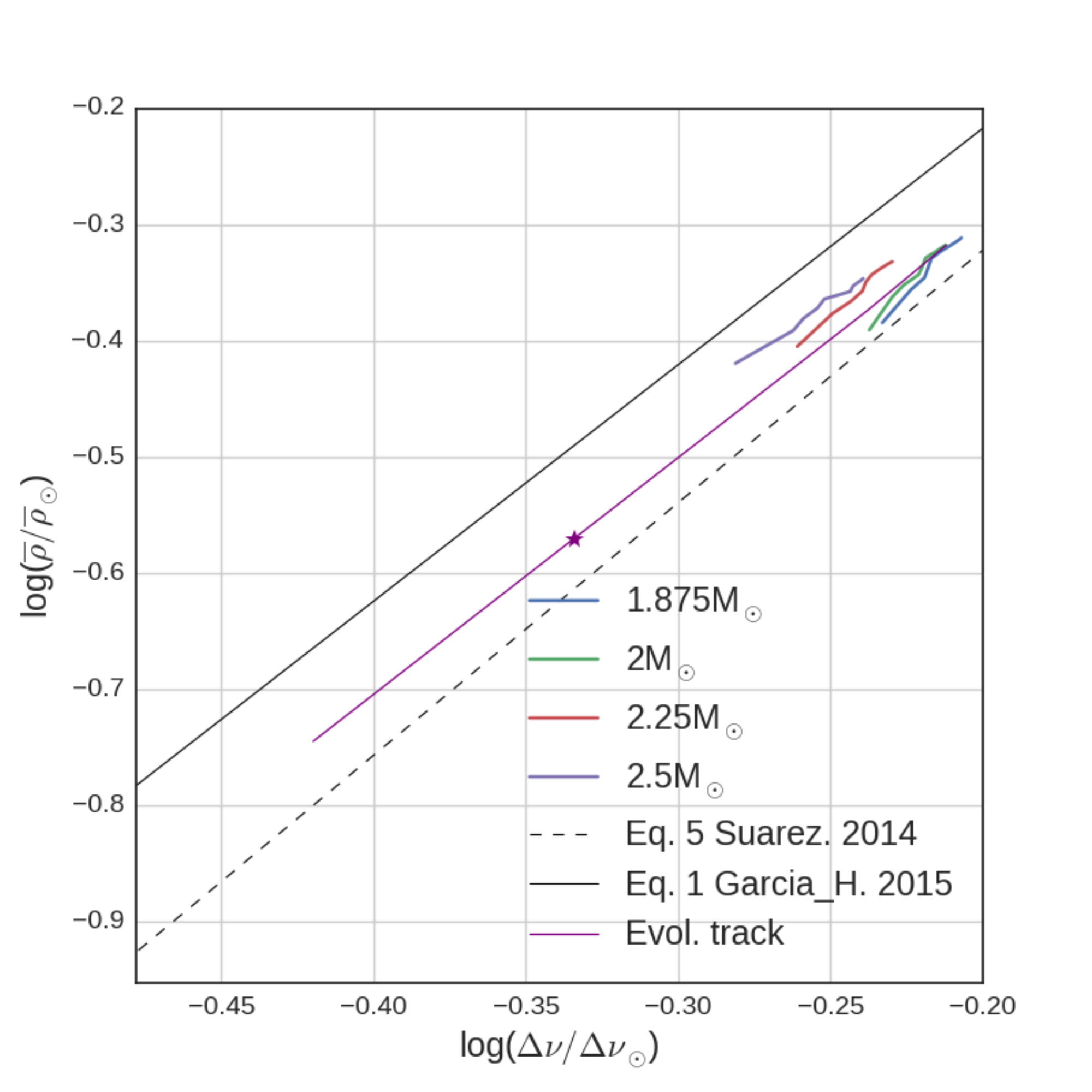}
  \caption{Relation between $\Delta \nu$ and $\sqrt{\overline{\rho}}$ for the rotating models considered. Also plotted are the relationships found by previous studies for comparison. The purple line shows how this relationship looks if the non-rotating model with 2$M_{\odot}$ is evolved and the star symbol along this line shows the mean density of a 2$M_{\odot}$ ZAMS model rotating at critical rotation. A colour version of this figure can be found online.}
  \label{fig:garcia}
\end{figure}

We compare the relation between $\Delta \nu$ and $\sqrt{\overline{\rho}}$ as a function of mass for the same evolution state, rotation, and the relation found by \cite{Garcia2015} and by \cite{Suarez2014} in Figure \ref{fig:garcia}.  Like Garcia Hernandez et al., we computed $\Delta \nu$ at sufficiently high $n$ so that we could avoid our second frequency region where the $p$ and $g$ modes overlap and the density of modes is comparatively high. This was done by locating the highest frequency mode with $g$ mode characteristics in our most rapidly rotating models and considering the next six modes for each $\ell$ that can be traced back to modes with $0 \leq \ell \leq 5$ in the non-rotating model. This gives us the modes with $3 \leq n \leq 8$ for which we compute $\Delta \nu$ averaged over the five large separations for each of the six values of $\ell$. The modes we use have sufficiently large $n$ that the variation of $\Delta \nu$ with $n$ and $\ell$ is relatively small in the narrow range we use. We do note that, for lower radial order $p$ modes that we do not include, the power law is nearly the same but the coefficient is different in the relation between the mean density and large separation for different modes.

We find the relation between $\Delta \nu$ and $\sqrt{\overline{\rho}}$ for increasing rotation is very similar to the relation found by \cite{Suarez2014} and \cite{Garcia2015}. In all cases, the slope is very similar to that of the classical period-root mean density relation.  The change with mass for the ZAMS models is not as steep, suggesting that the mean density is primarily an indication of evolution stage for models in this range. However, the correlation may not be unique when the results of a rapidly rotating model are compared with those of a slightly evolved non-rotating star, as is shown in Figure \ref{fig:garcia}. It goes without saying that obtaining a useful estimate of the mass depends crucially on the quality of the surface radius information.

We have also examined how individual frequencies relate to the mean density. \cite{Lovekin2009} found that a ``mean density'' computed using values of the surface radius at specific colatitudes gave a nearly linear relationship between the period and mean density when the radius defined at a colatitude of about 40 degrees was used. No particular significance was attached to this result. We have explored this further and agree with \cite{Lovekin2009} for the mode they used, but find that the colatitude required depends on the specific mode being examined. Specifically, we find that either larger $n$ or larger $\ell$ requires the colatitude for the surface radius to be used in the ``mean density'' to increase (i.e., the radius must increase). We attach no physical significance to this except as an expression that models with constant mass and evolutionary state but changing rotation do not constitute a homologous family.

\section{FINAL CONSIDERATIONS \label{sec5}}
We have examined the variation of both $g$ mode and $p$ mode oscillation frequencies in intermediate mass, uniformly rotating, ZAMS stellar models. The behavior of the $p$ mode frequencies with rotation is similar to that found for a 10$_{\odot}$ model by \cite{Lovekin2008}, suggesting that the behavior will be similar for models with masses in between.  By comparing models of different masses with the same surface shape, we found that the period-root mean density relation applies to the $p$ modes for rotating models to the same extent and with the same caveats as it does to non-rotating models. 

The $g$ mode frequencies are much less affected by rotation because their frequencies are determined in the region of the model between the convective core boundary and the lower boundary of the second helium ionization region. The region closer to the convective core is probably more important, but the entire region contributes to the frequency determination.  For sufficiently low frequencies, the frequency ratio of a given mode for two masses is almost independent of the mode. For these modes there is a slight trend of increasing frequency ratio with decreasing frequency as shown in Figure \ref{fig:freq-ratio}. 

The most complex region of the frequency spectrum is the region of overlap of the $p$ and $g$ modes. Because the $g$ mode frequencies increase with rotation and the $p$ mode frequencies decrease, the possibility of avoided crossings is high and the frequency ratio for specific modes between two masses may vary significantly as a function of rotation rate.  Figure \ref{fig:freq-ratio} suggests that modes with higher latitudinal variation show this feature more strongly than modes with lower latitudinal variation.

Another situation in which the scaling with mass will fail is when a given mode at one mass has undergone an avoided crossing while it has not at the other mass. If the modes rather than the shape of the latitudinal variation can be identified, the scaling holds. Once the mode has undergone the avoided crossings for both masses, the scaling is restored. Despite these problems, the scaling relationships found here should give general guidance on how to alter a model to obtain a better approximation to an observed mode frequency spectrum.

There remains the issue of matching modes observed in stars to modes computed for specific models. We have shown that there are certain scaling relations for the frequencies of a given mode from one model to another which may be helpful as part of this process. However, there are many gaps to fill before we can routinely associate a given observed mode with a specific computed mode. First and foremost is the need to have some reasonable estimate of the inclination of the rotation axis to the observer. If the inclination is low, one might use the shape of the wings of line profiles as for Vega (\citealt{Gulliver1991,Takeda2008,Yoon2010}). Stars with very high observed values of $V$ sin $i$ are presumably seen at high inclinations. Best is the determination of the inclination by optical interferometry, so far restricted to relatively few stars. Even with a reasonably determined inclination, the task is still formidable. An examination of the echelle diagram comparing observed mode frequencies with computed mode frequencies (e.g., Figure 5 of \cite{Deupree2011b} for $\alpha$ Oph) provides some indication of how far we have yet to go. Clearly, the density of computed modes requires a windowing process to help us associate an observed mode with a computed mode. For stars with a sizeable number of observed modes, patterns in the observed frequencies may exist for comparison with patterns in the computed frequency spectrum, much as for the Sun.  Stars with only a very limited number of modes with sizeable amplitudes may have variations in the colour difference between maximum and minimum light in different filters or variations in the line profiles that could help constrain the possibilities.  Similarly, theoretical determination of which modes are pulsationally unstable or otherwise of such low amplitude as to be unobservable at the appropriate inclination would be helpful.


\clearpage{}
\section*{Acknowledgments}

This research was performed under the auspices of the Canada Research Chairs program, and the authors are grateful for its support. We also thank the referee for encouraging us to expand the breadth of the paper.

\bibliographystyle{mn2e}
\bibliography{bibliography}

\end{document}